\documentclass{aa}
\usepackage{natbib}
\bibpunct{(}{)}{;}{a}{}{,} 
\usepackage{amsmath,amssymb,color}
\usepackage{mathrsfs}
\usepackage{graphicx}
\definecolor{timurblue}{rgb}{.39,.58,.93}
\begin{document}

\title{TeV cosmic-ray proton and helium spectra in the myriad model}
\titlerunning{TeV cosmic-ray proton and helium spectra}

\date{Processed \today ; Received ; Accepted ;\\Preprint numbers LAPTh-031/12, IFT-UAM/CSIC-12-68, SNU/2012-001}

\author{G. Bernard \inst{1} 
\and T. Delahaye \inst{2} 
\and Y.-Y. Keum \inst{3,4}
\and W. Liu \inst{5}
\and P. Salati \inst{1}
\and R. Taillet \inst{1}}

\institute{LAPTh, Universit\'e de Savoie, CNRS. B.P. 110. Annecy-le-Vieux F-74941. France
\thanks{Laboratoire d'Annecy-le-Vieux de Physique Th\'eorique, UMR5108}
\and
Instituto de F\'isica Te\'orica UAM/CSIC, Universidad Aut\'onoma de
Madrid, Cantoblanco, 28049 Madrid, Spain
\and
BK-21 Research Group, Department of Physics, Seoul National
University, Seoul 157-742, Republic of Korea
\and
Institute of the Early Universe, Ewha Womans University, Seoul, Republic of Korea
\and
Cosmic Dark-Energy and Dark-Matter Research Group, National Astronomical Observatories, Chinese Academy of Science,
Beijing 100012, Peoples Republic of China
}

\abstract
{Recent measurements of cosmic ray proton and helium spectra show a
hardening above a few hundreds of GeV. This excess is hard to understand
in the framework of the conventional models of Galactic cosmic ray production
and propagation.}
{We propose here to explain this anomaly by the presence of local
  sources (myriad model).}
{Cosmic ray propagation is described as a diffusion process taking place
inside a two-zone magnetic halo. We calculate the proton and helium fluxes
at the Earth between 50~GeV and 100~TeV. Improving over a similar analysis,
we consistently derive these fluxes by taking into account both local and
remote sources for which a unique injection rate is assumed.}
{We find cosmic ray propagation parameters compatible with B/C measurements
  and for which the proton and helium
spectra remarkably agree with the PAMELA and CREAM measurements over four
decades in energy.}
%


\keywords{
CREAM-PAMELA Anomaly,
Galactic Cosmic Ray Propagation,
Proton and Helium Spectra}

\maketitle
%
\section{Introduction}

\subsection{Observations of the CR proton and helium anomaly}

The energy spectrum of primary cosmic rays approximately behaves as $E^{-2.7}$,
in the 10~GeV to 100~TeV range. This is rather well understood if one assumes
that these cosmic rays are accelerated by energetic events such as supernova
explosion shocks, distributed evenly in the disk of our Galaxy. Once injected
inside the Galactic magnetic halo with a rate $q \propto {\cal R}^{-\alpha}$
where ${\cal R} \equiv {p}/{Ze}$ stands for the rigidity and
$\alpha \simeq 2.15 \pm 0.15$, particles are subsequently scattered by the
turbulent irregularities of the Galactic magnetic field. Their transport is
described phenomenologically by space diffusion with a coefficient
$K \propto {\cal R}^{\delta}$ whose energy dependence is characterized by the
index $\delta$. The boron to carbon (B/C) ratio is a tracer of cosmic ray (CR)
propagation and points towards a value of $\delta \in [0.4 , 0.85]$ for the
diffusion index. At high energy, the flux of a given primary CR species at the
Earth is given by $\Phi \propto {q}/{K}$ and falls with energy typically like
${1}/{E^{\, \alpha+\delta}}$. Its spectrum exhibits a power-law behavior whose
index is the sum of $\alpha$ and $\delta$.

However, this is challenged by the recent measurements of the absolute
high-energy CR proton and helium spectra that have been recently reported by
CREAM~\citep{CREAM2010,CREAM2011} and PAMELA~\citep{PAMELA2011} experiments.
Observations indicate the presence of an excess in the CR proton and helium
fluxes above 250~GeV/nuc. The single power-law hypothesis is rejected at 95 \% C.L.
The hardening of the proton spectrum occurs at $232^{+35}_{-30}$ GeV with a change
of the spectral index from $2.85 \pm 0.015 \pm 0.004$ to $2.67 \pm 0.03 \pm 0.05$.
For the helium data, the spectral index varies from $2.766 \pm 0.01 \pm 0.027$ to
$2.477 \pm 0.06 \pm 0.03$ with the hardening setting in at $243^{+27}_{-31}$ GeV/nuc.

\subsection{Current explanations}

Explanations of this anomaly have been tentatively given since its discovery.
They mainly imply a modification of the energy behavior of either the injection
spectrum $q(E)$ or the diffusion coefficient $K(E)$.
To commence, a break in $\alpha$ could arise from a modification of the conventional
diffusive shock acceleration (DSA) scheme, as suggested by~\citet{Malkov2012} and
\citet{Ohira2011}.
In the same vein, the possibility of different classes of CR sources has been proposed
some time ago by \citet{Stanev1993} and \citet{Zatsepin2006}. For instance, cosmic rays
accelerated in the magnetized winds of exploding Wolf-Rayet and red supergiant stars
could have a double spectrum, with a hard component produced in the polar cap regions
of these objects. According to \citet{Biermann2010}, this hard component would take over
the smooth one above a few hundreds of GeV, hence the observed break in the proton and
helium spectra.
Not so different is the proposition of~\citet{Yuan2011} where a spread in the injection
index $\alpha$ is introduced.
Another direction implies a modification of the diffusion coefficient $K$.
As proposed by~\citet{Ave2009}, the proton and helium anomaly could be due to
a welcome, but unexpected, decrease of the spectral index $\delta$ at high energy.
Recently~\citet{Blasi2012} have given some theoretical motivations to such changes in
diffusion.
A local variation of $K$ could also have a similar effect as suggested
by~\citet{Tomassetti2012}.
Finally, inspired by~\citet{Horandel2007}, \citet{Blasi2011} have invoked an unusually
strong spallation of the CR species on the Galactic gas.
This possibility has been recently criticized in a detailed analysis carried out by
\citet{Vladimirov2012} of some of the above mentioned solutions to the CR proton and
helium anomaly.

\subsection{Goal of the article}

In this paper, we compute the proton and helium spectra within the usual
framework or diffusive propagation, and using propagation parameters consistent
with B/C spectra, while taking into account the known local sources of cosmic
rays. We show that the proton and helium spectral hardening above 250 GeV/nuc
can be attributed to local sources of cosmic rays. These local sources are
associated to known supernova remnants (SNR) and pulsars, that can be found
in astronomical catalogs such as the Green catalog~\citep{Green2009} which can be
completed with the ATNF pulsar database~\citep{Manchester:2004bp}.
In our approach, there is no need to modify the conventional CR propagation model.
In particular, the variations with CR energy of the injection rate $q$ of individual
sources and of the space diffusion coefficient $K$ are power laws respectively
characterized by the spectral indices $\alpha$ and $\delta$.
This idea has already been suggested recently by \citet{Erlykin2011} who
explain the hardening with very few sources (mainly Monogem Ring) and by
\citet{Thoudam2012} who consider a catalog of 10 nearby sources. The principal
weakness of these analyses is the lack of a consistent treatment of the CR
spectra in the entire energy range extending from tens of GeV up to a few PeV.
This is particularly clear in~\citet{Thoudam2012} where the proton and helium
anomaly is derived from a handful of local sources, whereas the low energy
spectra of these species are not calculated but merely fitted in order to get
a value for $\alpha$ once $\delta$ has been chosen. It should be noted that
the magnitude of the CR proton (helium) flux is related over the entire energy
range to the injection rate $q$ of individual sources. The low energy (power-law
regime) and high energy (spectral hardening) parts of the CR spectra are
connected with each other. A consistent treatment of the problem requires that
the proton and helium fluxes are calculated over the entire energy range.
%
%
A crucial problem is also to understand why just a few local sources could
explain the spectral hardening at high energies whereas the bulk of the
Galactic sources is required in order to account for the power-law behavior
of the fluxes below 250 GeV/nuc. This aspect, which is not addressed
in the above mentioned analyses, bears upon the more general question of the
discreet nature of the sources. In the conventional model of CR propagation,
these are treated as a jelly spreading over the Galactic disk and continuously
accelerating cosmic rays. The question arises then to understand why
and in which 
conditions that scheme breaks down at high energies where local and discrete
objects come into play. The results presented here are based on a detailed
investigation of that question. \citet{Bernard2012} have recently shown how to
reconcile the presence of discrete sources with the conventional description
of CR production and propagation. We briefly recall the salient features of
their analysis.

\section{The proton and helium anomaly in the light of local sources.}

\subsection{Cosmic ray propagation with discreet sources.}

%
Once accelerated by the sources that lie within the Galactic disk, CR nuclei
diffuse on the irregularities of the Galactic magnetic field. The diffusion
coefficient $K = K_{0} \, \beta \, {\cal R}^{\delta}$ accounts for that process,
where $K_{0}$ is a normalization constant and $\beta$ denotes the particle
velocity. The magnetic halo (MH), inside which cosmic rays propagate before
escaping into intergalactic space, is assumed to be a flat cylindrical domain
which matches the circular structure of the Milky Way. The Galactic disk is
sandwiched between two confinement layers whose thickness $L$ is unknown and
turns out to be crucial in our investigation.
Stellar winds combine to generate a Galactic convection that wipes
cosmic rays away from the disk, with velocity
$V_{c}(z) = V_{c} \; {\rm sign}(z)$.
CR nuclei also undergo collisions with the interstellar medium (ISM)
with a rate
\begin{equation}
\Gamma_{\rm \! sp} = \beta \,
\left(
\sigma_{p{\rm H}} \, n_{\rm H} \, + \, \sigma_{p{\rm He}} \, n_{\rm He}
\right) \;\; ,
\end{equation}
where the densities $n_{\rm H}$ and $n_{\rm He}$ have been respectively
averaged to $0.9$ and $0.1$ cm$^{-3}$. The total proton-proton cross section
$\sigma_{p{\rm H}}$ has been parameterized according to \citet{Nakamura:2010zzi}
while $\sigma_{p{\rm He}}$ is related to $\sigma_{p{\rm H}}$ by the
\citet{Norbury:2006hp} scaling factor $4^{2.2/3}$. Similar scaling factors
have been used in order to derive the CR helium nuclei collision cross
sections from the proton case.
Above a few GeV, diffusive re-acceleration and energy losses may be
disregarded and the master equation for the space and energy number density
$\psi \equiv {dn}/{dT}$ of a given CR species simplifies into the diffusion
equation
\begin{equation}
{\displaystyle \frac{\partial \psi}{\partial t}} \, + \,
\partial_{z} (V_{c} \, \psi) \, - \,
K(E) \, \triangle \psi \, + \,
\Gamma_{\rm \! sp} \, \psi \, = \, q_{\rm acc} \;\; .
\label{master_equation}
\end{equation}
The CR transport parameters $K_{0}$, $\delta$, $L$ and $V_{c}$ can be
constrained from the B/C ratio -- see for instance
\citet{Maurin2001,Putze2010,Maurin2010}.
Our solution of the master equation~(\ref{master_equation}) is based
on the existence of a Green function $\mathcal{G}$ and is well adapted
to the presence of discreet and extended CR sources. As an illustration,
the CR proton density at the Earth may be expressed as the convolution
over space and time of the Green function $\mathcal{G}_{\rm p}$ with
$q_{\rm acc}$
\begin{equation}
\psi(\mathbf{x},t) =
{\displaystyle \int_{- \infty}^{\, t}} \!\!\! dt_{S}
{\displaystyle \int_{\rm MH}} \!\! d^{3}\mathbf{x}_{S} \;
\mathcal{G}_{\rm p} \left(
\mathbf{x} , t \, \leftarrow \, \mathbf{x}_{S} , t_{S}
\right) \,
q_{\rm acc}(\mathbf{x}_{S},t_{S}) \;\; ,
\label{psi_G_p}
\end{equation}
where $q_{\rm acc}(\mathbf{x}_{S},t_{S})$ is the CR proton injection
rate at the source located at $\mathbf{x}_{S}$ and at time $t_{S}$.
The propagator $\mathcal{G}_{\rm p}$ translates the probability for
a CR proton injected at position $\mathbf{x}_{S} \equiv (x_{S},y_{S},z_{S})$
and time $t_{S}$ to travel through the Galactic magnetic fields until it
reaches, at time $t$, an observer located at $\mathbf{x} \equiv (x,y,z)$.

%
In the conventional approach, the CR source term $q_{\rm acc}$ is
a continuous function of space and time. Steady-state is moreover assumed.
This is an oversimplification insofar as CR sources are actually discrete,
with an average supernova explosion rate $\nu$ of $0.8$ to $3$ events per
century \citep{Diehl:2006cf}.
In the stochastic treatment developed by \citet{Bernard2012}, sources
are modelled as point-like objects and
the production rate of CR
nuclei through acceleration is given by
\begin{equation}
q_{\rm acc}(\mathbf{x}_{S},t_{S}) \, = \,
{\displaystyle \sum_{i \in \cal{P}}} \,\, q_i \,\,
\delta^{3}(\mathbf{x}_{S} - \mathbf{x}_{i}) \,
\delta(t_{S} - t_{i}) \;\; ,
\label{q_acc_discret}
\end{equation}
where each source $i$ that belongs to the population $\cal{P}$ contributes
a factor $q_{i}$ at position $\mathbf{x}_{i}$ and time $t_{i}$. The total
flux $\Phi \equiv ({1}/{4 \pi}) \, \beta \, \psi$ at the Earth depends on the
precise locations and ages of all the sources and varies from one particular
population $\cal{P}$ to another. Because we do not know the actual
distribution of the Galactic sources that have generated the observed CR
flux, we must rely on a statistical analysis and consider the position and
age of each source as random variables. The CR flux $\Phi(E)$ at a given
energy $E$ behaves as a stochastic variable whose probability distribution
function $p(\Phi)$ has been studied in~\citet{Bernard2012}.
The conventional CR model is recovered by taking the statistical average
of the flux over the ensemble of all possible populations $\cal{P}$. This
average flux $\bar{\Phi}$ turns out to be the solution of
Eq.~(\ref{master_equation}) with a continuous source term $q_{\rm acc}$.
More exciting is the spread of the flux $\Phi$ around its average value
$\bar{\Phi}$. Using a Monte Carlo approach, \citet{Bernard2012} have shown
that if the magnetic halo is thin, the statistical fluctuations of the flux
may be significant. The residence time $\tau_{\rm dif} \sim {L^{2}}/{K}$ of
CR nuclei within the magnetic halo decreases with its thickness $L$. The number
$N$ of sources that contribute to the signal at the Earth scales as
$\nu \, \tau_{\rm dif}$. The smaller $L$, the smaller $N$ and the larger
the flux variance. Should the magnetic halo be sufficiently thin, we expect
fluctuations of the flux, especially at high energies where $K$ becomes
large. In this case, the
hardening of the proton and helium spectra appears to be a mere fluctuation
of the CR flux whose probability to occur is not vanishingly small.
According to this line of reasoning, the proton and helium anomaly results
from the particular configuration of the actual CR sources. These objects
are incidentally known in the nearby region for which catalogs of SNR and
pulsars are available.
The domain extending 2~kpc around the Earth and encompassing objects that
have exploded less than 30,000 years ago is defined as the local region.
The catalogs are no longer complete outside and fail to be reliable.
In the conventional CR model, the local sources would yield an average
contribution $\bar{\Phi}_{\rm loc}$ whereas the actual objects yield a
much larger flux $\Phi_{\rm cat}$. Denoting by $\Phi_{\rm ext}$ the flux
from the other sources, we infer a total signal at the Earth
\begin{equation}
\Phi \, = \, \Phi_{\rm cat} \, + \, \Phi_{\rm ext} \; ,
\end{equation}
to be compared to the prediction of the conventional steady-state model
%
\begin{equation}
\bar{\Phi} \, = \, \bar{\Phi}_{\rm loc} \, + \, \bar{\Phi}_{\rm ext} \;\; .
\end{equation}
The flux produced by the external sources has a very small variance as shown
in~\citet{Bernard2012}. We may then identify $\Phi_{\rm ext}$ with its statistical
average $\bar{\Phi}_{\rm ext}$.

%
\begin{table*}[t!]
\begin{center}
\begin{tabular}{|ccccccc|}
\hline
model 
& $K_0$ [kpc$^2$/yr] 
& $\delta$
& $L$ [kpc] 
& $V_{c}$ [kpc/yr] 
& $q_{\rm p}^{0}$  $\left[  {\rm GeV}^{-1}      \right]$ 
& $q_{\rm He}^{0}$ $\left[ ({\rm GeV/n})^{-1} \right]$ \\
\hline
A & $2.4\times 10^{-9}$   &  0.85 & 1.5  & 
$1.38\times 10^{-8}$  & $1.17 \times 10^{52}$ &
$3.22 \times 10^{51}$   \\
B & $2.4\times 10^{-9}$   &  0.85 & 1.5  & 
$1.38\times 10^{-8}$  & $0.53 \times 10^{52}$ &
$1.06 \times 10^{51}$  \\
MED & $1.12 \times 10^{-9}$ & 0.7  & 4   & $1.23\times 10^{-8}$   &
$15.8 \times 10^{51}$ & $3.14 \times 10^{51}$ \\
\hline
model 
& $\alpha_\text{p} + \delta$ 
& $\alpha_\text{He} + \delta$ 
& $\nu$ [century$^{-1}$]  
& H injection 
& He injection
& $\chi^2/\text{dof}$   \\
\hline
A &  2.9 & 2.8 & 0.8 & 0.19 & 0.05 & 0.61\\
B &2.85 & 2.7 & 1.4 & 0.12 & 0.07 & 1.09\\
MED &  2.85 & 2.7 & 0.8 & 0.148 & 0.07 & 1.3\\
\hline
\end{tabular} 
\end{center}
\caption{Sets of CR injection and propagation parameters discussed in
  the text.}
\label{table:parameters}
\end{table*}
%

\subsection{Scan of the parameter space}

%
The CR propagation parameters giving a good agreement with the secondary to
primary B/C ratio measurements have been determined by \citet{Maurin2001}
using the same propagation model as the one described above. For each of
these 1,600 different sets of parameters, CR propagation is specified by
$K_{0}$, $\delta$, $V_{c}$ and $L$.
The injection rate of CR species $j$ is assumed to be generically of the
form
\begin{equation}
q_{j}(p) = q_{j}^{0} \;
\left( {\displaystyle \frac{p}{1 \; {\rm GeV/nuc}}}
\right)^{\displaystyle - {\alpha}_{j}}
\;\; ,
\end{equation}
and to be the same for all CR sources. They are specified by the parameters
$q_{\rm p}^{0}$, $q_{\rm He}^{0}$, $\alpha_{\rm p}$ and $\alpha_{\rm He}$.
The final ingredient is the average supernova explosion rate $\nu$.
We have used these parameters to compute the proton and helium fluxes
$\bar{\Phi}_{\rm ext} + \Phi_{\rm cat}$ over an energy range extending from
50~GeV/nuc to 100~TeV/nuc, in order to compare with the PAMELA~\citep{PAMELA2011}
and CREAM~\citep{CREAM2010} data. At high energy, solar modulation has little
effect on the CR flux and it has not been taken into account in this study.
The quality of the fit to the data is gauged by the proton and
helium chi-squares $\chi^{2}_{\rm p}$ and $\chi^{2}_{\rm He}$. Our calculation
of the local contribution $\Phi_{\rm cat}$ to the flux is based on the Green
catalog~\citep{Green2009} and on the ATNF pulsar
database~\citep{Manchester:2004bp}.

%
We have performed a scan over the CR propagation parameters derived by
\citet{Maurin2001}. For a given set of these CR parameters, we have adjusted
the source parameters $q_{\rm p}^{0}$, $q_{\rm He}^{0}$, $\alpha_{\rm p}$,
$\alpha_{\rm He}$ and $\nu$ in order to get the lowest value for the total
chi-square $\chi^{2}_{\rm p} + \chi^{2}_{\rm He}$.
We first set the injection indices $\alpha_{\rm p}$ and $\alpha_{\rm He}$,
as well as the explosion rate $\nu$, and determine the best-fit values for
the injection normalizations $q_{\rm p}^{0}$ and $q_{\rm He}^{0}$. We then
vary $\nu$ from 0.5 to 3.5 $\text{century}^{-1}$ and the injection indices
$\alpha_{\rm p}$ and $\alpha_{\rm He}$ from 1.75 to 2.35 in order to get
the best adjustment of the source parameters to the CREAM and PAMELA data.
The injection indices $\alpha_{\rm p}$ and $\alpha_{\rm He}$ are determined
independently from each other. Observations point towards slightly different
power laws for the proton and helium fluxes at energies below
250~GeV/nuc.

\subsection{Results}

The range of spectra obtained by the scan described above is very
large, some of them are much higher and other much lower than the
measured spectra, for protons and helium. However, we find that some
of the parameter sets compatible with B/C give a good agreement with
measurements. Model A yields the best $\chi^2$, the resulting spectra are shown in Fig.~1,
and the corresponding parameters are reproduced in Table~1. 
The CR parameters of model B are the same as for model A. The average supernova
rate has been fixed to a value of 1.4 per century. The resulting spectra are shown in Fig.~2,
Model C corresponds to the MED configuration as defined by \citet{Donato:2003xg}
and best fits the B/C ratio. The half-thickness $L$ of the magnetic halo is
4~kpc. The resulting spectra are shown in Fig.~3. The agreement with
the CREAM and PAMELA proton and helium excess is still reasonable.
These sets of CR injection and propagation parameters featured in this
table have been shown
in~\citet{Maurin2001} to be compatible with the B/C ratio, and provide
reasonable to very good fits to the PAMELA and CREAM data from 
50~GeV/nuc to 100~TeV/nuc.
The proton and helium fluxes are simultaneously adjusted with the same values
of $K_{0}$, $\delta$, $L$ and $V_{c}$. The injection indices $\alpha_{\rm p}$
and $\alpha_{\rm He}$ are determined independently from each other. The average
supernova explosion rate per century is denoted by $\nu$. The results of the fits
to the proton and helium spectra are gauged by the total reduced chi-square
$\chi^{2}_{\rm red}$ (see Table~1).

%
\begin{figure}[t!]
\includegraphics[width=\columnwidth]{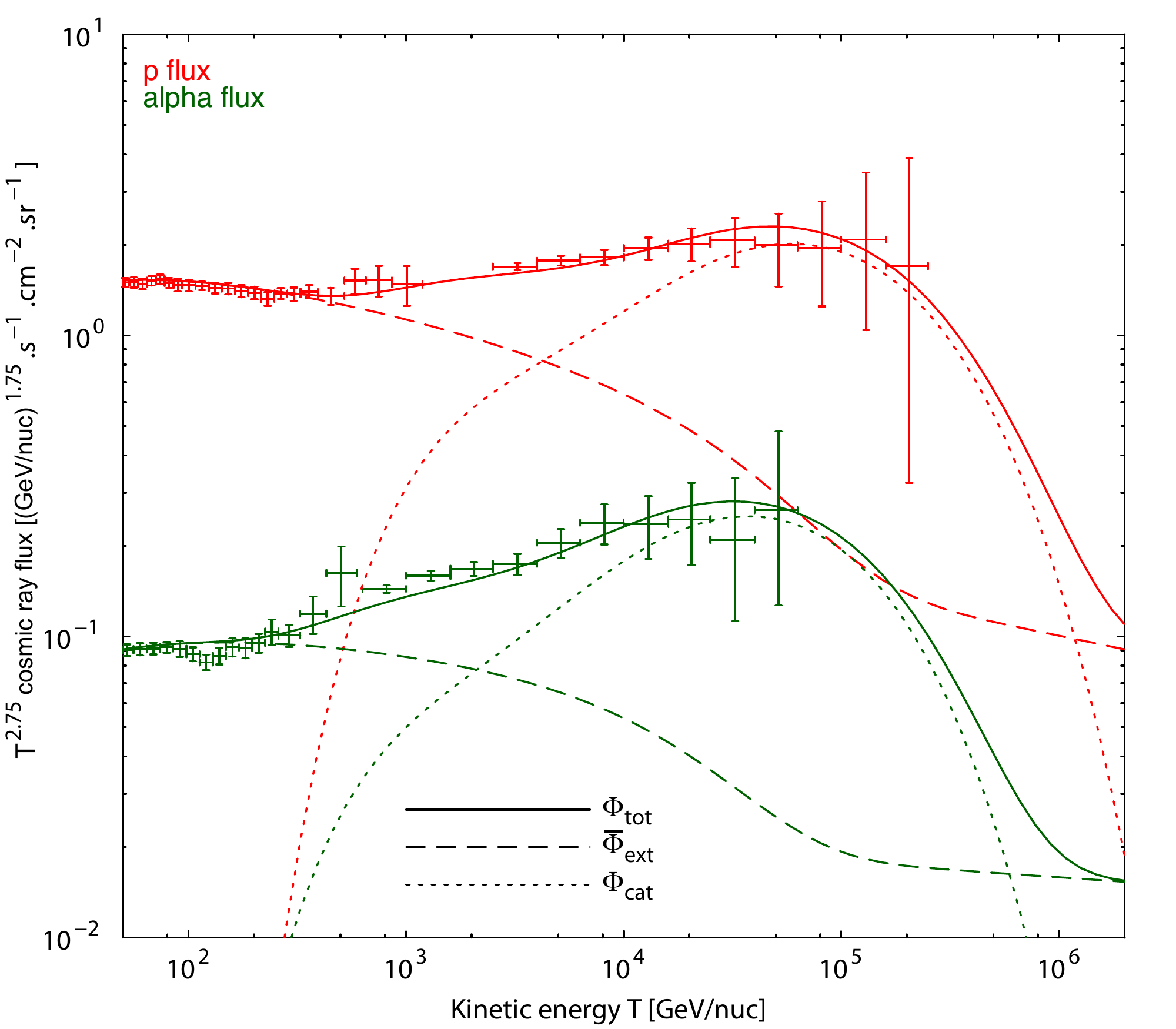}
\caption{
Proton (upper curve) and helium (lower curve) spectra in the range extending
from 50~GeV/nuc to 100~TeV/nuc, for the propagation parameters of
model A (see Table~1), giving
the best fit to the PAMELA~\citep{PAMELA2011} and
CREAM~\citep{CREAM2010} data : supernovae explosion rate $\nu =
0.8 \;\text{century}^{-1}$. Solid lines show the total flux,
short-dashed lines show the flux due to the sources of the catalog,
and the long-dashed curve the flux due to the rest of the sources.}
\label{Fig1} 
\end{figure}
%

%
\begin{figure}[t!]
\includegraphics[width=\columnwidth]{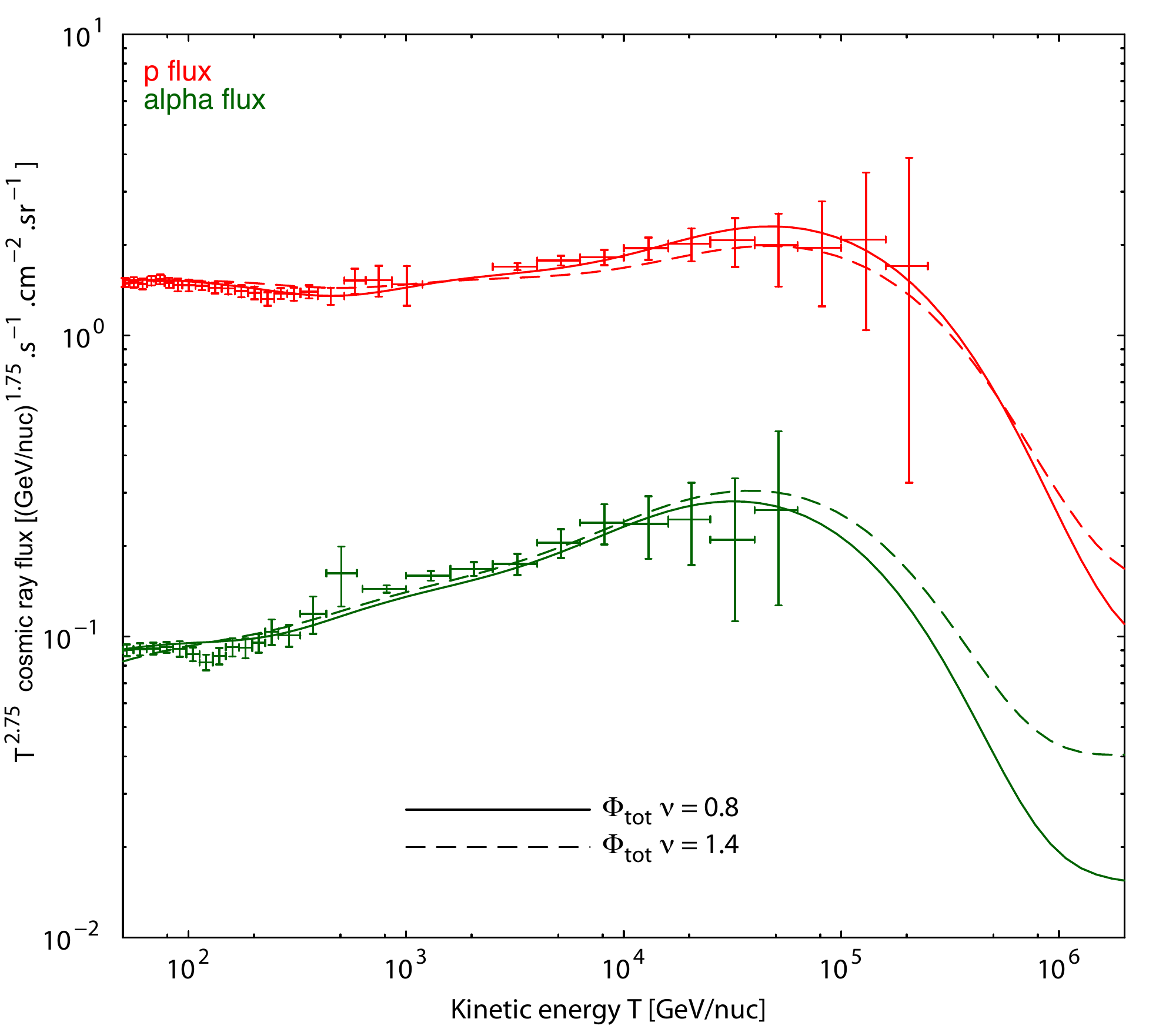}
\caption{
Same as previous figures for models A and B (see Table~1), for two values of the supernovae explosion
rate,  $\nu = 0.8 \;\text{century}^{-1}$ and $\nu = 1.4 \;\text{century}^{-1}$.}
\label{Fig3} 
\end{figure}
%

%
\begin{figure}[t!]
\includegraphics[width=\columnwidth]{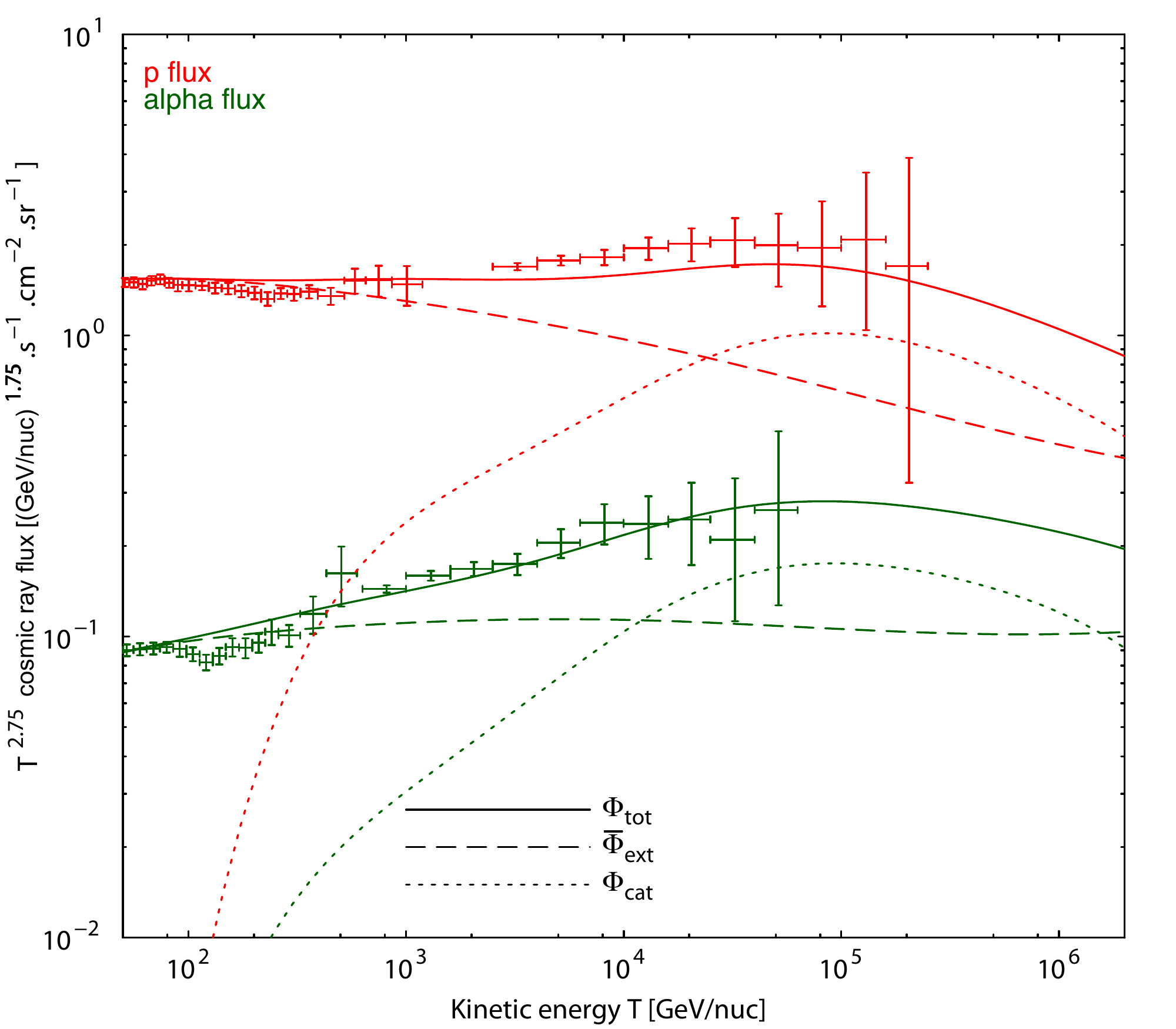}
\caption{
Same as before, for the MED propagation parameters (see Table~1).}
\label{Fig2} 
\end{figure}
%

\section{Discussion of the results.}

This excellent agreement makes us confident that the proton and helium anomaly
can actually be explained by existing local sources which have been extracted
from SNR and pulsar surveys. The model which we have presented here is quite
simple. Refining it is beyond the scope of this paper. Some directions can
nevertheless be given in order to improve the solution which we have just
sketched.
To commence, the best fits are obtained for a rather small value of the magnetic
halo thickness $L$. This trend can be understood as follows. 
As already explained, the thinner the magnetic halo,
the smaller the number $N$ of sources which contribute to the total signal and
the larger the injection rate $q$ of individual sources. The contributions
$\Phi_{\rm cat}$ and $\bar{\Phi}_{\rm loc}$ from the local region are no longer
swamped in the total flux when $L$ is small. This may be a problem as recent
studies~\citep{Strong2010,Bringmann2012} of the gamma-ray and synchrotron diffuse
emissions seem to favour rather large values of $L$. A possible improvement of
our model would be to distribute the CR sources within spiral arms and to take
into account the rotation of the Galaxy. The sources which we have considered
here in order to derive the contribution $\bar{\Phi}_{\rm ext}$ to the flux
are equally spread along the azimuthal direction. It would be interesting to
investigate if $\bar{\Phi}_{\rm ext}$ decreases in a more realistic setup.
Notice also that in order to get a significant injection rate $q$, we are naturally
driven towards a small supernova explosion rate $\nu$. The values found for models
A and B are close to 1 explosion per century, at the lower edge of the plausible
range, although not excluded \citep{2006Natur.439...45D}.


The local sources which we have extracted from the catalogs correspond to
a larger rate $\nu$ of $3.3$ events per century. Because the Sun lies near two
Galactic arms, the average explosion rate in our neighborhood could reasonably be
higher than the mean rate of the Galaxy. This is actually supported by our catalog.
As shown in Fig.~7 of~\citet{Bernard2012}, the number of known sources in our
vicinity is compatible with a value of $\nu$ larger than 3 explosions per century
for the past $3 \times 10^{4}$ years (depending on the radial distribution of the
CR sources along the Galactic disk). Taking into account the Galactic spiral arms
and their rotation could lead to a larger value of the average explosion rate and
alleviate the apparent discrepancy between the average and local values of $\nu$.
Finally, we have modeled the supernova explosions as point-like events. Cosmic
rays are believed to be accelerated in the shocks which follow these explosions
and which propagate in the ISM during $10^{5}$ years. The injection sites are more
spherical shells than points. Depending on the CR energy, \citet{Thoudam2012}
quote escape times between $500$ and $10^{5}$ years after the stellar explosion.
The injection takes place from a remnant whose radius varies from 5 to 100~pc.
Taking into account the actual structure of CR accelerators could substantially
modify the contribution $\Phi_{\rm cat}$ to the total signal, allowing larger values
of $L$ to provide acceptable fits to the PAMELA and CREAM data.
The simplistic solution to the proton and helium anomaly which we have sketched 
in this paper is definitely promising in spite of the above mentioned problems
and should motivate further investigations.

\section{Conclusions.}

Taking into account the discreteness of cosmic rays sources in the
solar neihbourhood, we found that the proton and helium spectra
computed in a diffusion model agree with
  the PAMELA and CREAM measurements over four decades in energy, for
  some cosmic ray propagation parameters which are also compatible with B/C measurements
Even if the excess at high energy happened not to been confirmed by
further measurements and analysis, the proton and helium spectra could
be used to put severe constraints on the 
parameters describing the diffusion of cosmics rays in our Galaxy.
We expect that the study of the anisotropy induced by the discreteness
of the sources will also provide very valuable information on the
source distribution, as well as on the propagation parameters.

%

\begin{acknowledgements}
This work was supported by the Spanish MICINNs Consolider-Ingenio 2010 Programme
under grant CPAN CSD2007-00042. TD also acknowledges the support of the MICINN under
grant FPA2009-08958, the Community of Madrid under grant HEPHACOS S2009/ESP-1473, and
the European Union under the Marie Curie-ITN program PITN-GA-2009-237920.
YYK would like to thank LAPTh at Annecy in France for the warm hospitality he has
received during his visits there. His work is partially supported by the Research
Professorship (A) of Ewha Womans University during 2010-2011, and by the Basic Science
Research Program through the National Research Foundation of Korea (NRF) funded by the
Korean Ministry of Education, Science and Technology (Grant No:2009-0090848).
WL would like to thank people in LAPTh for their hospitality during over one year and
partial financial support from LAPTh. He would like to thank Xuelei Chen, Youjun Lu and
Xiaoxia Zhang for helpful discussions. He is grateful to the China Scholarship Council
for its financial support (No. 2009491052).
PS expresses his gratitude to the Institut universitaire de France (IUF) for academically
and financially supporting his research.
\end{acknowledgements}

%
\bibliographystyle{aa}
\bibliography{PandA_PRL}
%
\end{document}